\documentclass[oldversion]{aa}
\usepackage{graphicx}
\usepackage{natbib}
\usepackage{txfonts}

\begin{document}

   \title{Numerical determination of the material properties of porous dust
     cakes.}

\titlerunning{Properties of porous aggregates}
\authorrunning{D. Paszun \& C. Dominik}

\author{ Paszun D.  \thanks{\emph{Present address:} Anton Pannekoek Institiute,
University of Amsterdam, Kruislaan 403, 1098SJ Amsterdam The Netherlands} \and
Dominik C.}

\offprints{D. Paszun, \email{dpaszun@science.uva.nl}}

\institute{Astronomical Institute "Anton Pannekoek", University of Amsterdam,
Kruislaan 403 1098 SJ Amsterdam}

\date{Received $<$date$>$ / Accepted $<$date$>$}

\abstract {The formation of planetesimals requires the growth of dust particles
  through collisions.  Micron-sized particles must grow by many orders of
  magnitude in mass.  In order to understand and model the processes during this
  growth, the mechanical properties, and the interaction cross sections of
  aggregates with surrounding gas must be well understood.  Recent advances in
  experimental (laboratory) studies now provide the background for pushing
  numerical aggregate models onto a new level.  We present the calibration of a
  previously tested model of aggregate dynamics.  We use plastic deformation of
  surface asperities as the physical model to bring critical velocities for
  sticking into accordance with experimental results.  The modified code is
  then used to compute compression strength and the velocity of sound in the
  aggregate at different densities.  We compare these predictions with
  experimental results and conclude that the new code is capable of studying the
  properties of small aggregates.}


\maketitle


\section{Introduction}
It is commonly accepted that planets form in protoplanetary disks. Giant planets
are believed to be produced via one of two competing scenarios.  The first one
is a gravitational instability \citep{2007ApJ...661L..73B}. Clumps of gas
(intermixed with dust) may become gravitationally bound and contract forming
giant planets.  This model is hotly debated
\citep[i.e.][]{2000ApJ...529.1034P,2006ApJ...651..517B,2007ApJ...656L..89B,
  2002Sci...298.1756M,2004ApJ...609.1045M,2007ApJ...661L..77M}. The second,
commonly accepted scenario is known as the core accretion model
\citep{1996Icar..124...62P}. In this mechanism the solid core of a planet forms
first and gas is accreted onto that core later.  The core itself is believed to
form by collisional accumulation of planetesimals. Terrestial planets, due to
their similarity to the giant planets' cores, must form out of planetesimals
with help of gravitational interactions. The formation of planetesimals is also
a subject of debate. The scenario initially proposed by
\citet{1973ApJ...183.1051G} assumed a laminar structure of the disk. When the
dust sublayer reaches a critical thickness, a gravitational instability may form
planetesimals from the dense and gravitationally bound concentrations of
dust. However, this mechanism requires an extremely laminar nebula. The shearing
motion of the dust layer causes a Kelvin-Helmholtz instability and therefore
limits the thickness of the sublayer. The gravitational instability of the dust
layer is then prevented. This was proved by \citet{1984Icar...60..553W},
\citet{1993Icar..106..102C} and \citet{2006ApJ...643.1219J}.

The core accretion model requires to grow micron sized dust grains into km sized
planetesimals.  These are over 27 orders of magnitude in mass. Radial drift
alone may move dust particles inwards onto the central star before they reach
bigger size. The drift can move meter sized particles all the way in within only
100 orbits \citep{1977MNRAS.180...57W}.  Relative velocities of large bodies set
another barrier against accumulation of large particles
\citep{1993prpl.conf.1031W,2005PhRvE..71b1304W,2007A&A...466..413O}. A recent
study by \citet{2006ApJ...636.1121J} showed however, that planetesimals might be
produced by a gravitational collapse in the presence of turbulence. Meter sized
boulders concentrated in high pressure turbulent eddies in the disk may form
gravitationally bound clumps.  Even if this process really can be made to work,
dust particles must first grow over $18$ orders of magnitude in mass to be able
to form that dense accumulations. This growth must happen through the
collisional sticking of dust particles.

The initially small, micron sized, dust grains (further referred to also as
\emph{monomers}) collide and stick to each other due to attractive surface
forces \citep{1971ProcRSocLonA..324.301J}. The fine dust is very well coupled to
the gas, meaning that the particles move collectively. The relative velocities,
due to the Brownian motion \citep{1828PhMa..4.161B}, are very small. These
conditions lead to a quasi-monodisperse growth that preferentially collides
particles of \emph{similar} size. The aggregates (further referred to also as
agglomerates or particles) formed in this case are fractal
\citep{1999Icar..141..388K,2000PhRvL..85.2426B,2004PhRvL..93b1103K,2006Icar..182..274P}.

The Brownian growth produces aggregates with very open structure and the fractal
dimension $D_\mathrm{f}=1.5$
\citep{2000PhRvL..85.2426B,2004PhRvL..93b1103K,2006Icar..182..274P}.  However,
the gas density influences the growth. When the collisions are no longer
ballistic, due to high gas density and thus short stopping length, the fractal
dimension decreases and in the case of a very high gas density it may even
approach unity \citep{2006Icar..182..274P}. This process may play a role in the
inner most regions of protoplanetary disks.

The subsequent growth of the dust aggregates eventually leads to decoupling of
the dust component from the gas. The particles start to be affected by
turbulence, sedimentation and radial drift
\citep{1977MNRAS.180...57W,1980Icar...44..172W,1984Icar...60..553W}. The
relative velocities thus increase and collisions occur preferentially between
agglomerates of \emph{different} sizes \citep{1993prpl.conf.1031W}. When the
collision energy becomes higher than some threshold restructuring energy,
aggregates are compacted and their fractal dimension approaches $D_\mathrm{f}=3$
\citep{1997ApJ...480..647D,2000Icar..143..138B}. Further growth and compaction
causes the particles to decouple from the gas more effectively and the relative
velocities increase further. \citet{1997ApJ...480..647D} and
\citet{2000Icar..143..138B} showed that the outcome of collisions can be
categorized in terms of collision energy. Perfect sticking without restructuring
occurs when the collision energy is lower than the threshold rolling energy,
which is the energy needed to roll a single particle over an angle of $\pi/2$.
When the collision energy reaches this limit, the aggregates start to compress
upon impact. The maximum compaction occurs when the collision energy equals the
rolling energy times the number of contacts in the aggregates. The particles
start to lose monomers, when the impact energy reaches the number of contacts
in the aggregate times the energy needed to break one contact. Catastrophic
disruption occurs when the energy reaches several times the number of contacts
times the breaking energy. Therefore, as particles grow and decouple from the
gas, eventually relative velocities are reached that will lead to shattering of
the colliding aggregates \citep{2007A&A...466..413O}.

This general picture has several missing elements.  Although numerous
experiments have been performed in different size and velocity regimes
\citep[for a review see Blum \& Wurm, ARAA,2008 in press]{2004PhRvL..93b1103K,
  2000Icar..143..138B, 2005PhRvE..71b1304W, 2005Icar..178..253W}, our
understanding of the processes involved in the growth of aggregates is still
incomplete. The restructuring mechanism for instance is understood only
qualitatively. The degree of the collisional compaction still remains a
mystery. This, however, is crucial for the growth of the meter sized
aggregates. The sticking efficiency as a function of the particle density along
with the density evolution must be determined in order to understand
quantitatively the growth of meter sized boulders.

The low strength of the aggregates due to the fractal structure leads to
fragmentation in the case of fast impacts. The distribution of fragment sizes
must be also determined as a function of the collision energy. This makes it
possible to keep track of the realistic size distribution in the disk. Small
particles, if not replenished in the disk, are very quickly swept up by larger
grains \citep{2005A&A...434..971D}. Thus the small particles should be supplied
to the disk by some processes. \citet{2007InPress..DD07} show that the infall of
grains from the parent cloud is rather unrealistic and requires fine tuning of
parameters to reproduce observational data. Thus collisional fragmentation is
the most likely mechanism that can explain the population of small grains in
protoplanetary disks. However, long before fractal aggregates approach
velocities high enough to cause fragmentation, they undergo collisional
compression \citep{2000Icar..143..138B,1993prpl.conf.1031W}. Thus the only
scenario leading to fragmentation of fractals is collision with much larger
non--fractal particle.

The current understanding of processes like sticking, bouncing and fragmentation
of aggregates is poor. The understanding of this processes on micro scales may
fill these gaps and allow extrapolations to the larger sizes.  This may be
resolved by two approaches. Experiments may be performed in a laboratory to
provide useful data. However, this way is available only for aggregates below
centimeter in size. Larger particles are currently inaccessible for experiments.
The second, theoretical approach, understanding the material properties of
porous matter reach aggregates of meter size and beyond. Thus it would be ideal
to provide theoretical predictions for centimeter sized and larger aggregates.

\citet{2004Icar..167..431S} developed a model capable to simulate big -- meter
sized and larger aggregates, using Smoothed Particles Hydrodynamics (SPH). In
this case one particle in the model is an aggregate characterized by material
properties like, compressive strength, tensile strength, density and sound
speed. In order to obtain some of these properties he fitted power-law functions
to experimental data of compression and tensile strength. 

This method was then also used by \citet{2007A&A...470..733S}. Further studies
require, however, a more precise determination of the material properties of
porous bodies. These may be obtained in laboratory experiments or in computer
simulations as presented in this work.

More realistic Solar Nebula dust analogs were used in experiments by
\citet{2006ApJ...652.1768B}.  They investigated aggregates made of micron sized
dust particles. Moreover different aggregates used in these experiments
consisted of spherical or irregularly shaped monomers. The compression and
tensile strength curves for this dust cakes were determined.

Although experiments give a quantitative description of processes in cm sized
aggregates, it is difficult to access for small, a few microns size aggregates.
Aggregate dynamics models \citep{1997ApJ...480..647D,2002Icar..157..173D} are
ideal for simulations of these small scale structures. This method spatially
resolves single monomers, which is certainly needed to understand physics of
bigger -- meter sized bodies. Until now the main drawback of this aggregate
dynamics model was missing quantitative agreement with experiments, even though
the qualitative agreement has already been established
\citep{2000Icar..143..138B}. Thus a calibration of this model is required.

Another aggregate dynamics model was recently presented by
\citet{2007ApJ...661..320W}. They made use of potential energies in order to
derive forces acting between grains in different degrees of freedom. In this
case they present just a 2D case, but their results are in agreement with
findings of N--body model presented earlier by \citet{1997ApJ...480..647D}. Wada
et al. (ApJ,2007; submitted) shows the first results of compression and
disruption of 3 -- dimensional aggregates in head -- on collisions. Although
the model is qualitatively in agreement with experiments, as is studied by
\citet{1997ApJ...480..647D}, the quantitative mismatch is still present.
\citet{2007ApJ...661..320W} do not show any solution or workaround to the
quantitative disagreement between theory and laboratory experiments.  The work
presented here addresses this issue and provides mechanisms that fit our model
to the empirical data.

In this paper we present the calibration of the aggregates dynamics model
developed by \citet{1997ApJ...480..647D} and \citet{2002Icar..157..173D}. We fit
the code using experimental data and further test it. Modifications that are
implemented in the code are presented together with a few possible application of
the model.

\section{The model}
In order to study the agglomeration mechanisms involved in the growth of
planetesimals, we use the SAND code (Soft Aggregate Numerical Dynamics)
developed by \citet{2002Icar..157..173D}. It is a N--body model of a system of
spherical particles interacting via surface forces. 

Two monomers feel the attractive force only when they are in contact. The
surfaces of the particles deform and form a contact area. When the particles are
pulled outwards, increasing the relative distance, the area decreases and the
monomers are pulled back to the equilibrium position by the surface forces. The
compression of the system on the other hand leads to increase of the contact
area and repulsive force pushing the particles apart. A detailed description of
the surface forces was provided by \citet{1971ProcRSocLonA..324.301J} (further
referred to as JKR). The influence of adhesion forces on the contact between
particles was also studied by \citet{1975JournCollInterfSci...53..314D}. The
model is able to treat long range magnetic forces \citep{2002Icar..157..173D},
however these are not the subject of our current study.

There are two main processes that govern the events during a collision. The
first one is breaking a contact between two monomers. This dissipates part of
the energy and weakens or destroys aggregates participating in the
collision. The second process is a rolling motion of a monomer over another
grain. This also dissipates energy and causes a restructuring of an
aggregate. This restructuring may be attributed to both compression, which
results in strengthening the aggregate, or decompression, i.e. weakening the
aggregate's structure.

The JKR theory predicts a critical force that is needed to separate two
particles. This prediction was tested for the case of micron size Silica spheres
by \citet{1999PhRvL..83.3328H}. Two monomers were pulled off each other using an
Atomic Force Microscope. The measured pull-off force was in agreement with the
force predicted by the JKR theory. The pull-off force is
\begin{equation}
F_{\mathrm{c}} = 3 \pi \gamma R,
\end{equation}
where $\gamma$ is a surface energy and $R=\bigl( \frac{1}{R_1} + \frac{1}{R_2}
\bigr)^{-1}$ is a reduced radius of the spheres in contact. Thus the results of
the experiments confirm the theoretical predictions.

A similar experiment was performed in order to determine the horizontal forces
acting on the particles in contact (first studied theoretically
by\citet{1997ApJ...480..647D}). The horizontal displacement of the contact zone
causes a torque acting against the displacement. The resulting rolling friction
was measured in laboratory experiments, again using an AFM
\citep{1999PhRvL..83.3328H}.

\subsection{Rolling friction}
Rolling friction is one of the most important energy dissipation channels in
the restructuring of aggregates \citep{1997ApJ...480..647D}. It is thus of a
great importance to treat it in a correct way. The theoretical derivation by
\citet{1997ApJ...480..647D} shows that the energy associated with initiating
rolling is expressed as
\begin{equation}
e_\mathrm{roll}=6 \pi \gamma \xi_\mathrm{crit}^2,
\label{rolling_friction}
\end{equation}
where $\xi_\mathrm{crit}$ is a critical displacement at which the inelastic
behavior occurs and energy is dissipated.  $\xi_\mathrm{crit}$ was initially
assumed to be of the order of inter -- atomic distances
\citep{1997ApJ...480..647D}. The rolling motion causes shift of the contact
area, implying that at one end the contact has to be broken in order to form it
at the other side of the contact area. The experimentally determined friction
showed that the critical displacement must be larger. The determined value was
$\xi_\mathrm{crit}=3.2$~nm meaning that the displacement is rather of the order
of ten inter atomic distances. \citet{2002Icar..157..173D} took that already
into account and used in their model the value of
$\xi_\mathrm{crit}=10$~$\mathrm{\AA}$. This however was applied to different
material than the one investigated in the lab.  We followed
\citet{1999PhRvL..83.3328H} and applied larger value of
$\xi_\mathrm{crit}=20$~$\mathrm{\AA}$, which is approximately $10$ inter atomic
distances.

\subsection{The pull-off force and the critical velocity for sticking}
At low velocities the dominating mechanism that dissipates energy is rolling. At
higher impact speeds, however, other channels become more important.
\citet{1993ApJ...407..806C} and \citet{1997ApJ...480..647D} calculated how much
of the initial energy is dissipated in the collision between two particles. This
gives the maximum energy at which the particles can stick in a head--on
collision. The critical velocity is given by
\begin{equation}
v_\mathrm{crit} = 1.07 \frac{\gamma^{5/6}}{E^{*1/3}R^{5/6} \rho ^{1/2}},
\label{critical_velocity}
\end{equation}
where $\gamma$, $R$ and $\rho$ are surface energy, reduced radius and mass
density, respectively. $E^{*-1}= \frac{1-\nu_1^2}{E_1}+\frac{1-\nu_2^2}{E_2}$ is
a reduced elasticity modulus with $\nu_i$ and $E_i$ being Poisson's ratio and
Young's modulus, respectively, of grain $i$. When eq.~\ref{critical_velocity} is
applied to a $R_1=0.6$~$\mu$m silica grain, impacting flat silica surface, it
gives the critical velocity $v_\mathrm{crit}=0.18$~m$\mathrm{s}^{-1}$. This was
again tested in experiments. \citet{2000ApJ...533..454P} showed that such a
particle can stick to the surface at significantly higher velocities, of the
order of $1$~m$\mathrm{s}^{-1}$.  The measured velocities were
$1.2\,\pm\,0.1$~m$\mathrm{s}^{-1}$ for $0.6$~$\mu$m grains and
$1.9\,\pm\,0.4$~m$\mathrm{s}^{-1}$ for $0.25$~$\mu$m grains. They used slightly
different definition of the critical velocity. The critical velocity was defined
as the velocity at which the probabilities of sticking and bouncing are equal
$50\%$. They measured the sticking velocity for different materials, shapes and
sizes of particles. The resulting critical velocity was shown to depend on size
of the monomer as a power law with index of about $0.53$. Although the
theoretically derived slope $R^{-5/6}$ of the power law is different from the
empirical data, they are still consistent within error bars. Moreover the power
law was fitted to only two data points.

We will assume that the theoretical dependence on the radius is correct and
consistent with the experiment. The discrepancy between the absolute values of
the critical velocity however have to be revised, doing so is absolutely
essential for meaningful results.

In order to better understand the mechanisms involved in the sticking of the
monomers we refer to experiments by \citet{2000ApJ...533..454P} and
\citet{1975JournCollInterfSci..51.58D}. They investigated the bouncing of micron
sized spheres off a flat silica surface. In the former experiment the impacting
particle was made of silica, while the latter one used soft polystyrene grains.

\citet{1975JournCollInterfSci..51.58D} determined experimentally the critical
velocity to be about $1$~ms$^{-1}$. The ratio of the rebound velocity to the
incident speed (coefficient of restitution) decreases in this case with
decreasing impact velocity. At very large velocities of about $20$~ms$^{-1}$
\citet{1975JournCollInterfSci..51.58D} noticed a decrease of the restitution
coefficient with increasing impact speed. He related that behavior to plastic
deformation.

In the experiment by \citet{2000ApJ...533..454P} the first positive value of the
coefficient of restitution indicates a critical velocity of the order of $\sim
1$~ms$^{-1}$.

In order to scale the model to the experimental data we introduce a mechanism
that dissipates part of the initial energy at the moment of first contact.  The
plastic deformation of small asperities on the surface of the grain is an easy
way to dissipate the energy and increase the critical velocity. Below we will
estimate the central pressure in order to compare that to the yield strength of
$104$~MPa \citep{2000msea.book.....C} for silica\footnote{Since the yield
strength of brittle material like silica is not defined, flexular strength is
given.}. \citet{1993ApJ...407..806C} on the other hand argue that the yield
strength for very small bodies is of the order of $0.2$ times the Young's
modulus of the material. This corresponds to $~11$~GPa for silica.

The maximum pressure in the contact area occurs in the center of the contact.
The radial distribution of pressure in the contact zone is the following
\citep{1996PhMA...73.1279D}
\begin{equation}
p(r,a) = 6\frac{F_\mathrm{c}}{\pi a_0^2} \frac{a}{a_0} (1 - \Bigl( \frac{r}{a}
\Bigr) ^2)^{1/2} - 2 \frac{F_\mathrm{c}}{\pi a_0^2} \biggl( \frac{a}{a_0}
\biggr)^{-1/2} (1 - \Bigl( \frac{r}{a} \Bigr) ^2)^{-1/2}.
\label{eq::pressure_distribution}
\end{equation}
$a_0$ in this equation is the equilibrium contact radius given as
\begin{equation}
a_0 = \biggl( \frac{9 \pi \gamma R^2}{E^*} \biggr)^{1/3}.
\label{eq::a_0}
\end{equation}
Now we need to estimate the maximum contact radius that
is reached during the collision. The radius of the contact area increases with
increasing normal load. The force applied to the sphere during the collision
is $F=d(mv)/dt$, where $d(mv)$ is the momentum of the impacting particle and
$dt\approx 10^{-9}$~s is the collision time. With this force applied, the
contact radius is
\begin{equation}
a = \biggl( \frac{3R}{4E^*} (F+6\pi \gamma R + \sqrt{(6\pi\gamma
R)^2 + 12\pi\gamma RF} )\biggr)^{1/3},
\label{eq::a}
\end{equation}
as derived by \citet{1971ProcRSocLonA..324.301J}.  Thus in the case of the
velocity $v=1$~m/s, the central pressure is of the order of $1$~GPa. This
pressure however is exactly in between the two values of the yield strength
specified above, making the plastic deformation possible. The lower limit of the
yield strength ($104$~MPa) is reached at radii $r/a<0.914$, which makes $83\%$
of the contact area exposed to potential plastic deformation.

This idea was investigated before by \citet{1990AreosolSciTech..12.497T}. The
energy consumption in such a deformation can be expressed as
\begin{equation}
E_\mathrm{asp} = Y V_\mathrm{asp},
\label{plastic_energy}
\end{equation}
where $Y$ is the yield strength of the deforming material and $V_\mathrm{asp}$
is the volume of the asperities that are flattened during the collision. To
calculate the dissipated energy we need to get the volume of asperities.  We
follow \citet{1990AreosolSciTech..12.497T} again and calculate the maximum
contact area given by
\begin{equation}
a_\mathrm{max}=\Biggl( \biggl( \frac{3R}{4E^*}\biggr)[F_\mathrm{eq}+6\pi \gamma
R +\sqrt{(6\pi \gamma R)^2+12\pi \gamma R F_\mathrm{eq}}]\Biggr)^{1/3}
\label{contact_radius}
\end{equation}
\citep{1993ApJ...407..806C} and the equivalent impact force given by
\citet{1990AreosolSciTech..12.497T} as
\begin{equation}
F_\mathrm{eq} = 2.5^{3/5} E_\mathrm{kin}^{3/5} R^{1/5}E^{*\,2/5}.
\label{equivalent_force}
\end{equation}
Now we can get the volume of the deformed material by multiplying the contact
area by the volume of a single asperity, and the number of asperities per
surface area
\begin{equation}
V_\mathrm{asp} = \frac{2}{3} \pi r_\mathrm{asp}^3 N_\mathrm{asp} * \pi
a_\mathrm{max}^2,
\label{volume}
\end{equation}
where $r_\mathrm{asp}$ is the radius of a single asperity. Here we assume the
bumps to be hemispheres distributed homogeneously over the surface of a particle
or a target. In our case we use a parameter that describes the total volume
plastically deformed per unit of the surface area. This way we have only one
parameter that defined how efficiently the energy is dissipated. Note that in
reality the pressure in the contact area is compressive in the center and
tensile on the edge. Thus our parameter may be slightly higher than what we
present.

\begin{figure}[!h]
\resizebox{\hsize}{!}{\includegraphics{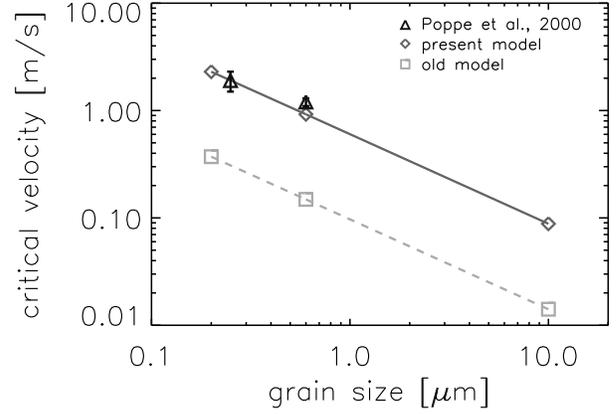}}
\caption{Critical velocity as a function of a grain size. Squares indicate the
model without additional dissipation process, diamonds show present model with
the energy dissipation by plastic deformation of surface roughness and triangles
show experimentally determined values by \citet{2000ApJ...533..454P} with error
bars.  }
\label{fig::vcrit}
\end{figure}

Fig.~\ref{fig::vcrit} shows the critical velocities determined experimentally
\citep{2000ApJ...533..454P} for two different grain sizes. Also the critical
velocity obtained using our model is plotted showing that we successfully fitted
our model and we can reproduce the experimental results.

In our model the volume of the asperities deformed upon collision is very
small. When we assume surface roughness to be hemispheres with radii
$r_\mathrm{asp}=1$~nm, the fraction of the area occupied by the asperities is
only about $2\%$. We can thus still say that the molecular size asperities may
be responsible for increasing the critical velocity. Moreover we see in
fig.~\ref{fig::vcrit}, that the dependence of the sticking velocity on the grain
size is not affected and the only difference is that the energy regime has been
shifted to the level measured in experiments.

In order to avoid confusion, we must stress here again that we do not claim that
the plastic dissipation takes place during the collisions of silica spheres. We
simply implement additional scaling parameter in order to fit our model to the
experimental data. The discrepancy between the empirical data and theory should
be further investigated. We also think that plastic deformation might need
somewhat more attention, because of the very high pressure that is present in
the center of the contact area and relatively low strength of the material
considered in this work.

\subsection{Excitation and Cooling}
In parallel to the energy dissipation due to contact breaking, energy can be
dissipated via other channels in the vertical degree of freedom (along the line
connecting centers of two grains in contact). Two grains held together by the
surface force vibrate relative to each other
\citep{1995PhMA...72..783D}. Ideally the oscillation is frictionless and no
dissipation occurs. In reality, however, the vibration causes an oscillation in
the size of the contact area. When decreasing the contact size, part of the
energy is dissipated by breaking the connections at the edges of the contact
area. This ultimately leads to a cooling of the aggregate and damps the
vibrations. If this mechanism is not taken into account, successive slow
collisions may heat up the aggregate and ultimately lead to ``evaporation'' of
monomers from the aggregate surface.

In fact we observed this phenomenon in our simulations of linear
chains. Successive collisions of grains with an aggregate caused an increase of
the amplitude of the oscillation and eventually lead to breaking of several
contacts. All individual collision velocities were far below the sticking
velocity, proving that this mechanism may produce artificial results.

To resolve this problem we introduced a weak damping force, which was intended
to slowly dissipate the vibrational energy. The force is acting in the vertical
direction in respect to the contact area. The damping force is expressed as
\begin{equation}
F_\mathrm{damp}= \mathrm{const} \, v_z,
\label{damping_force}
\end{equation}
where $v_z$ is the vertical component of the relative velocity, and
$\mathrm{const}$ is arbitrarily chosen to damp exponentially 95 percent of the
vibration energy within $\sim 100$ oscillation periods.  By tuning the constant
to this value we made sure that the damping force influence is not significant
within short timescales of a few first vibration periods, where the sticking or
bouncing event is determined. In the first period the dissipation takes place
and if the particle looses not enough energy it will bounce. If the damping
force is too strong, it may remove enough energy to allow sticking. We made
sure, that the main energy dissipation process that sets the critical velocity
is the plastic deformation mechanism and the sticking velocity is not affected
by the presence of the damping force.

To show the energy leakage due to the damping force we calculated the total
energy in a vibrating pair of monomers. The particles were displaced from the
equilibrium position and the kinetic and potential energies were calculated. In
the case of a frictionless oscillation the potential energy is 
\begin{equation}
E_\mathrm{p} = \int_0^\delta F(\delta') \mathrm{d} \delta',
\label{eq::Epot_general}
\end{equation}
where $\delta$ is a displacement such that the distance between centers of the
two grains is $r_1+r_2-\delta$. The vertical force
\begin{equation}
F(a) = 4 F_\mathrm{c} \biggl( \biggl(\frac{a}{a_0} \biggr)^3 - \biggl(
\frac{a}{a_0} \biggr)^{3/2} \biggr)
\label{eq::F_of_a}
\end{equation}
depends on the contact size $a$ and equilibrium contact radius $a_0$. The
integral~(\ref{eq::Epot_general}) may be then changed into
\begin{equation}
E_\mathrm{p} = \int_0^{a} F(a') \mathrm{d} a'.
\label{eq::Epot_a}
\end{equation}
To get the energy we solve eq.~\ref{eq::Epot_a} by changing the variable from
$\delta$ to $a$ using the following relation
\begin{equation}
\delta = \frac{1}{2} \frac{a_0^2}{R} \biggl( 2 \biggl( \frac{a}{a_0} \biggr)^{2}
- \frac{4}{3} \biggl( \frac{a}{a_0} \biggr)^{1/2} \biggr).
\label{eq::delta_of_a}
\end{equation}
The potential energy is then given as
\begin{equation}
E_\mathrm{p} = 4 F_\mathrm{c} \frac{a_0^2}{R} \biggl[ \frac{2}{5} \biggl(
\frac{a}{a_0}\biggr)^5 - \frac{2}{3} \biggl(\frac{a}{a_0}\biggr)^{7/2} +
\frac{1}{6} \biggl(\frac{a}{a_0}\biggr)^2 \biggr].
\label{eq::Epot_vibra}
\end{equation}
The kinetic energy is simpler. We just add kinetic energies of all monomers
together
\begin{equation}
E_\mathrm{k} = \sum_i \frac{m_i v_i^2}{2},
\label{eq::Ek}
\end{equation}
where $m_i$ and $v_i$ are mass and absolute velocity of the $i$-th grain.

\begin{figure*}
\centering \includegraphics[width=17cm]{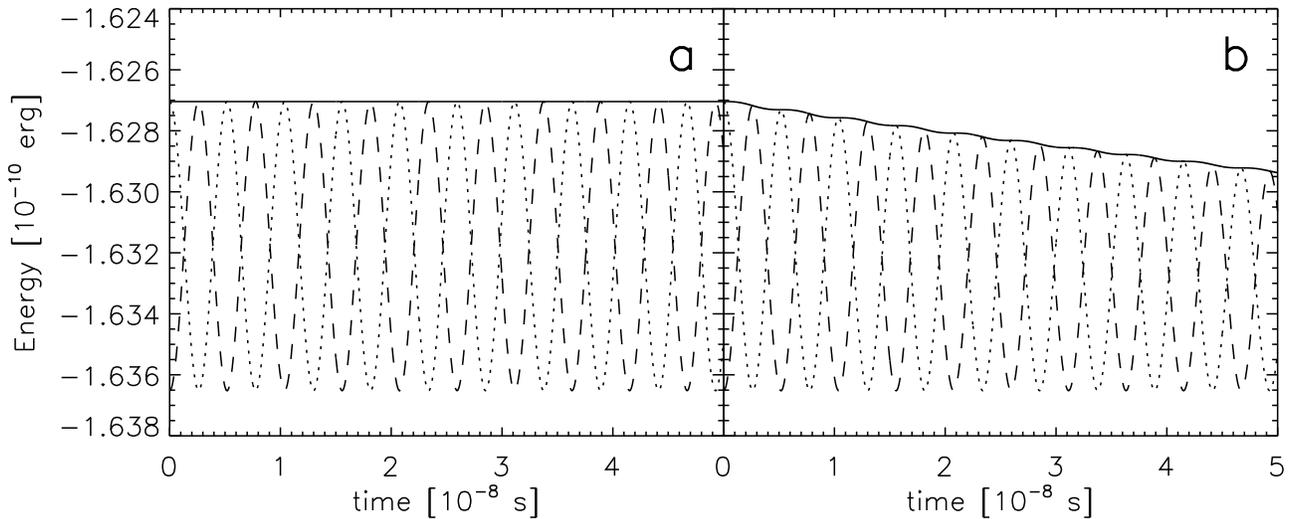}
\caption{The total energy is redistributed between the kinetic energy (dashed
line) and the potential energy (dotted line). At the maximum and minimum
separation the potential energy is maximum, while in the equilibrium position
the kinetic energy is the highest. The kinetic energy was shifted down for
better overview (see text). No damping case -- a, and the energy leak form the
system due to the damping force -- b.}
\label{fig::vibration}
\end{figure*}

The total energy $E_\mathrm{tot}$ in the case of a frictionless oscillation is
plotted in fig.\ref{fig::vibration} a. To better see variations in the kinetic
energy we shifted it to the level of the potential energy. The shift is equal to
the potential energy of the system in the equilibrium $E_\mathrm{p} =
\int_0^{\delta_0} F(\delta') \mathrm{d} \delta'$. The total energy in this case
is conserved and the amplitude is constant.
When we enable the damping force the energy starts to leak. The potential and
kinetic energies may be calculated using the same formulas. The energy
dissipation is presented in fig.~\ref{fig::vibration} b.

The energy loss due to the damping force is very weak and removes about $95\%$
of the total energy within approximately $100$ vibration periods. Thus other
processes like rolling, and breaking dominate the energy dissipation. In
particular the critical velocity for sticking is not influenced at all by this
damping force, because the assumed plastic deformation of asperities dissipates
nearly the entire collision energy within the first vibration period.

\section{Applications}
The presented model is a very good tool that can be used to study the
aggregation of dust particles. However, the modifications that we introduced in
the previous section need further verification. Although the critical velocity
is well fitted to the experiments, it is very useful to further test the model
against laboratory experiments. When the model is successfully verified, it can
be applied confidently in the study of aggregates dynamics. Compaction and
fragmentation of dust aggregates are processes that require further
investigation \citep{2005A&A...434..971D}. Our model is perfectly suited to this
task. Dynamical processes, involving small sub-mm sized aggregates, can be
studied in depth using our tool. We can provide detailed understanding of
micro-physics that governs compaction and fragmentation in this scale. Phenomena
involving aggregates of mm and larger sizes require an entirely different
approach.  They must be treated with different methods i.e. Smoothed Particles
Hydrodynamics (SPH). However, to do this, material properties, such as
compressive strength and sound speed, are needed. Below we present simulations
which can be directly compared to experiments and provide great test of our
model. We simulate compression of a dust cake and determine the compressive
strength of such cake. We also determine the sound speed in such a porous
aggregates.

\subsection{Compression}
\citet{2004PhRvL..93k5503B} measured the compressive strength of dust cakes in
the laboratory. The sample was prepared by random ballistic deposition (RBD).
Single monomers were shot from one direction at low velocity (hit--and--stick)
and grow the dust agglomerate. The resulting ``cake'' was about $2$~cm in
diameter and similar in height. The finished cake was later placed between two
flat surfaces. The load was applied onto the upper surface and caused it to move
towards the lower surface compressing the dust sample.  In order to simulate
compression of a dust cake we need to prepare adequate setup.

\subsubsection{Setup}
In the experiment by \citet{2004PhRvL..93k5503B} the applied pressure resulted
in compression of the dust cake. Measurement of the cake volume resulted in
determination of a filing factor $\phi$.

Our setup was organized in a similar manner to the experimental one. Our code,
however, can handle only spherical particles. Thus instead of two planes we used
two very big ``wall'' grains, each with diameter of $2 \cdot 10^{-2}$ cm. The
sample was shaped as a cylinder and the distance between two compressing
``wall'' grains was adjusted to exactly fit the micro dust cake.

Our dust cake was grown via particle cluster aggregation method (PCA). The
target aggregate (initially a monomer) was randomly oriented before each
collision with a grain approaching at a random impact parameter. Any successful
hit resulted in perfect sticking without any restructuring. The new target was
then randomly oriented, and a new grain was shot at it again. The resulting
aggregate was then shaped into a cylinder by removing all grains outside of the
desired contour.

The size of the cake was $13.2 \,\mu$m in diameter and it was $13.2\,\mu$m
high. When filled by $291$ monomers with radius of $0.6\,\mu$m the filling
factor of the cake was $\phi=0.146$. For comparison the dust cake used in real
experiment had the filling factor $\phi=0.15\pm0.01$. The monomer size in that
experiment was also of similar size $r_0=0.75\,\mu$m.

The compression was done by moving one of the large grain-surfaces at constant
velocity. While it was approaching, the second grain was fixed in its position
and was unable to move even upon extreme pressures. The sample was thus
compressed by one surface against the other one. The initial setup is presented
in fig.~\ref{fig::setup}a. Two large grains on both sides of the aggregate are
the back wall, fixed plane in the right and the approaching, compressing plane
in the left. The aggregate is placed in between and the particles can escape
sideways increasing this way the diameter of the cake.

In order to simulate quasi static compression we fixed the velocity of the
compressing ``wall'' grain to $0.05$ m/s. This is over an order of magnitude
lower than a critical velocity and much lower than the sound speed in this
medium (see section 3.2). Thus the assumption that we are in a quasi static
regime is reasonable. The dominant acceleration of a single monomer, in this
case, is due to surface forces.

For the purpose of this compression simulation, we disable the net force acting
onto the compressing ``wall'' particle, but we save the record of this force for
each time step. Thus the approaching surface cannot be stopped and moves with
constant speed. For each time step the net force is stored and later used to
determine a pressure.

At each successive step the dust cake becomes slightly more compressed. The
degree of compression can be related to the force that was needed to get the
cake into this state. 

\subsubsection{Results}
The small size of the dust cake used in this numerical experiment causes a low
number of contacts of the compressed aggregate with the approaching
surface. Consequently, the net force applied to the compressing surface was
strongly variable. Every new contact formed between the dust cake and the
incoming surface resulted in a sudden decrease of the force. The new connection
is initially stretched and thus the surface is attracted. Similarly, oscillations
of monomers at the surface cause additional variation. This makes it more
difficult to uniquely determine the compression force. To overcome this problem,
for ten each successive time steps we choose the one with the highest force,
because ultimately this is the force required to compress the cake.

The pressure was calculated as $P=\frac{F}{S}$, where $F$ is the normal load and
$S$ is the cross-section area. 
\begin{figure*}
\centering \includegraphics[width=17cm]{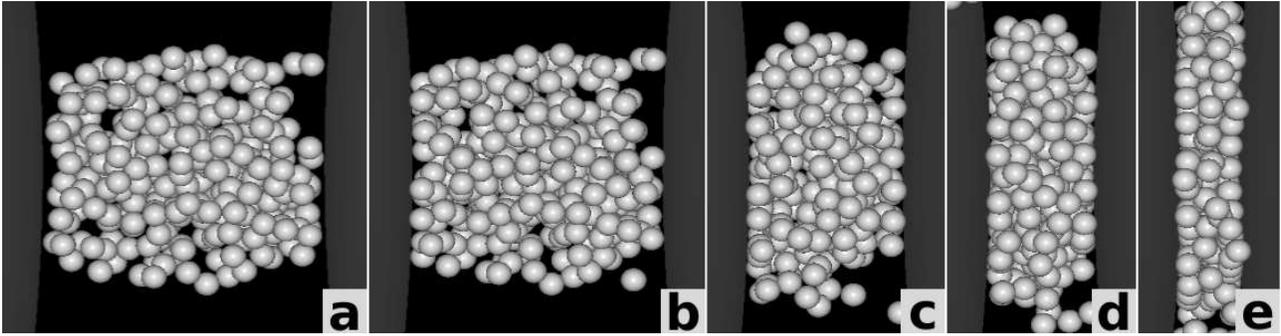}
\caption{The setup of the experiment. The dust cake in the center is compressed
  with different pressure.  Initial arrangement (a), results of compression at
  $2 \cdot 10^2$~Pa (b),$2 \cdot 10^3$~Pa (c), $5 \cdot 10^3$~Pa (d), $1 \cdot
  10^4$~Pa (e).}
\label{fig::setup}
\end{figure*}

\begin{figure}[!h]
\resizebox{\hsize}{!}{\includegraphics{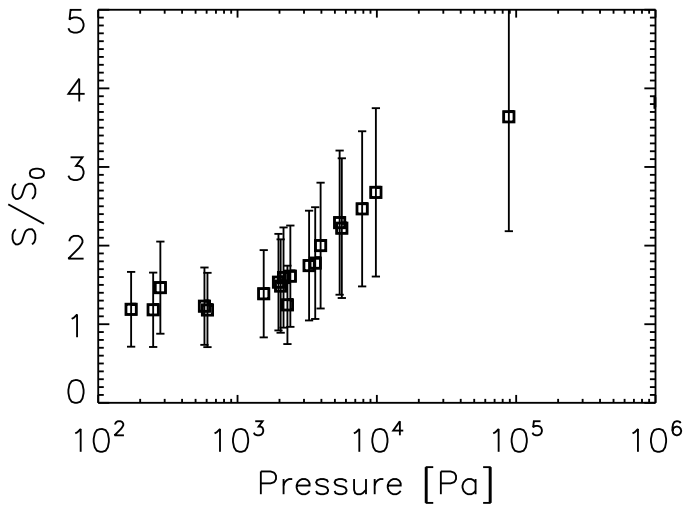}}
\caption{The ratio of the cross-section of the dust cake to the initial
  cross-section as a function of the applied compressive pressure. The data
  presented here is from a single experiment. Each point is plotted with a 40\%
  uncertainty in determination of the crosssection (see text).}
\label{fig::cakecrosssection}
\end{figure}

Fig.~\ref{fig::setup} shows the initial setup and the results of compression
with increasing pressure. The lowest pressure cannot restructure the aggregate.
The height of the dust cake is almost unchanged. A higher pressure of 2~kPa
compresses the cake. The monomers in the cake's surface are pushed down into the
cake. At a pressure of $1\cdot10^4$~Pa the cake is compressed significantly,
causing a horizontal flow of the particles and thus an increase of the cake
diameter. The evolution of the dust cake cross-section is shown in
fig.~\ref{fig::cakecrosssection}.
The relative increase is affected by the size of the dust cake. What may be a
negligible boundary effect in a large macroscopic aggregate, here it has a large
impact onto the entire dust cake. When the diameter of the cake increases just
by a diameter of a single monomer, the area increases by $40$ $\%$. In the
experiment by \citet{2004PhRvL..93k5503B} a cake of about $2$ cm diameter was
compressed and the final increase of the cross-section was measured to be larger
by a factor of $1.6$ relative to the initial cross-section. In our simulation
the area increases almost 4 times, subject to significant uncertainties in the
determination of the dust cake cross-section. The initially cylindrical shape
changed upon compression into an irregular profile.

\begin{figure}[!h]
\resizebox{\hsize}{!}{\includegraphics{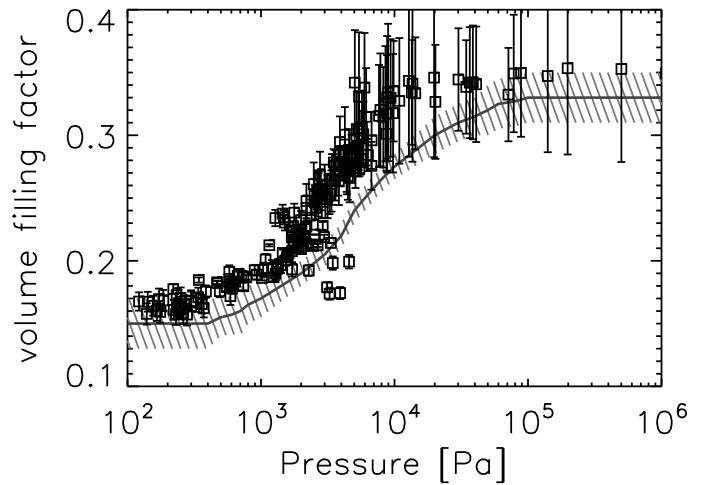}}
\caption{The volume filling factor vs normal pressure. The solid line indicates
  the results of the laboratory experiment by \citet{2004PhRvL..93k5503B}. The
  dashed area is the error to that data. Squares show results of our single
  simulations. Error bars are due to difficulty in determination of the volume
  of the final aggregate.}
\label{fig::compression}
\end{figure}

To determine the volume filing factor, we used only the inner part of the cake.
A cylindrical volume enclosing initially the dust aggregate, and used to
determine the filling factor, was decreased in height only. In this way we
reduced the uncertainty that arose from the boundary effects. The initially
porous aggregate is deformed and can expand. Particles are pushed into the cake,
filling voids. Thus the filling factor must increase as an effect of
compression. Fig.~\ref{fig::compression} shows our results together with the
data obtained in the laboratory experiments \citep{2004PhRvL..93k5503B}.
Low pressures are unable to affect the aggregate. However, the boundary effects
also cause problems. Initially the ``wall'' particle connects to the cake by
only one monomer. This causes very strong variations in pressure.  We show the
compression curve starting in the point when the ``wall'' particle has made $5$
contacts with the cake. At this point, however, a few grains are already pressed
into the dust cake, causing an increase of the volume filling factor. Thus our
compression curve in fig.~\ref{fig::compression} is shifted upwards. The filling
factor is also overestimated by using big spherical grains as the compressing
surfaces. The height of the dust cake we used is calculated as $D - 2
R_\mathrm{compress}$, where $D$ is distance between centers of the two big
grains and $R_\mathrm{compress}$ is their radius. Thus the volume occupied by the
dust cake is actually slightly larger. This effect is initially less important,
as it contributes only about $2$ $\%$ error. However, it gets more relevant at
larger pressure, where the cross-section of the dust cake is larger and the
volume smaller.

Error estimation was done using the central parts of the dust cake, where
boundary effects are smaller. The filling factor was determined by calculating
the volume of monomers enclosed in a cylindrical volume. The error was then
estimated to be the difference between the filling factors determined for two
different cylinders. The first one had a radius of $6$~$\mu$m, while the second
one was one monomer radius smaller.

The volume filing factor is initially constant, until the pressure reaches the
value of about $5 \cdot 10^2$ Pa. The filling factor increases until it reaches
the maximum compression of about $\phi=0.35$ and remains constant. The most
interesting result is the resemblance of our findings to the lab
experiments. The onset of compression fits the experimental data of
\citet{2004PhRvL..93k5503B}. Our compression curve follows the laboratory data
tightly ending up at a similar value of the filling factor. The small
differences between two curves are most likely due to small size of our
simulated dust cake.

\subsection{Sound speed}
One of the very important properties in the porous material is the sound
speed. It must be lower than the bulk sound speed because the mass is being
moved by a force acting on a very small contact area. The collision of two
grains with supersonic velocities can result in complete disruption of those
bodies. We will first derive an analytical formula for the sound speed. For this
we apply the JKR theory and assume that the signal transported is a very small
perturbation. The main assumption is that the two monomers in contact behave
like a perfect spring which is the case for small amplitudes. However, the
contact forces are asymmetric in respect to the equilibrium position. Thus
\emph{compression} of two particles results into different forces than
\emph{stretching} it to the same displacement. We can therefore expect that
large amplitudes lead to modified sound speeds. In the following sections we
present an analytical approach and later we show our simulations for both the
simplified case of a linear chain of monomers and the general case of a non
fractal aggregate.

\subsubsection{Analytical solution}
Every two monomers that are in contact are held together by surface forces
acting in the contact area \citep{1971ProcRSocLonA..324.301J}. Mutual attraction
inevitably leads to a vibrational spring-like motion, which in linear
approximation can be written as
\begin{equation}
F=-k(\delta_0 - \delta),
\label{springEQ}
\end{equation}
where $\delta$, $\delta_0$, $k$ and $F$ are displacements, equilibrium
displacement, spring constant and restoring force, respectively. Note that the
displacement $\delta$ is defined in a way that it increases, when the distance
between two monomers is decreasing. In our case we want to determine the spring
constant, which is necessary to find the sound
speed. \citet{1971ProcRSocLonA..324.301J} showed that the surface force in the
contact area is related to the contact area radius $a$ by
\begin{equation}
F=4F_\mathrm{c} \biggl( \biggl( \frac{a}{a_0} \biggr)^3 - \biggl( \frac{a}{a_0} 
\biggr)^{3/2} \biggr),
\label{F_of_a}
\end{equation}
We also know the relation between the displacement and contact radius
\begin{equation}
\delta = 6^{1/3} \delta _\mathrm{c} (2a_0^{-2}a^2 - 4/3a_0^{-1/2}a^{1/2}),
\label{delta_of_a}
\end{equation}
where $\delta_\mathrm{c}=1/2 \frac{a_0^2}{6^{1/3}R}$ is a critical displacement
for breaking the contact.  With these equations we can determine the spring
constant and later the sound speed.  First we differentiate eq.~\ref{F_of_a} and
\ref{delta_of_a} at $a=a_0$ to get
\begin{equation}
\frac{dF}{da}\Bigg|_{a=a_0}=\frac{6F_\mathrm{c}}{a_0}
\label{dF_of_a}
\end{equation}
and
\begin{equation}
\frac{\mathrm{d}\delta}{\mathrm{d}a}\Bigg|_{a=a_0}=\frac{5\delta_0}{a_0},
\label{ddelta_of_a}
\end{equation}
where $\delta_0=\frac{a_0^2}{3R}$ is an equilibrium displacement.  Now we can
write
\begin{equation}
\frac{\mathrm{d}F}{\mathrm{d}\delta}=\frac{\mathrm{d}F}{\mathrm{d}a}
\frac{\mathrm{d}a}{\mathrm{d}\delta}= \frac{\mathrm{d}F}{\mathrm{d}a}\Bigl(
\frac{\mathrm{d}\delta}{\mathrm{d}a} \Bigr)^{-1}
\label{dF_ddelta_da_ddelta}
\end{equation}
and substituting eq.~\ref{dF_of_a} and \ref{ddelta_of_a} we get
\begin{equation}
\mathrm{d}F=\frac{6}{5} (9\pi \gamma R^2 E^{\star 2})^{1/3} \mathrm{d}\delta .
\end{equation}
Since the restoring force is exactly opposite, we may add the minus sign here
and see that the spring constant $k$ is
\begin{equation}
k=\frac{6}{5} (9\pi \gamma R^2 E^{\star 2})^{1/3}.
\label{spring_const}
\end{equation}

With the spring constant of the oscillating system of two spheres in contact we
can proceed to the calculation of the sound velocity. For a spring the velocity
of sound is given by
\begin{equation}
c_\mathrm{s}=\sqrt{\frac{kL}{\rho _\mathrm{l}}},
\label{c_s}
\end{equation}
where $L$ is the total length of the spring (or set of the springs) and $\rho
_\mathrm{l}$ is a mass of the spring per unit length. We can then determine the
velocity in a linear chain of $n$ grains in contact, each $r_{0}$ in radius. The
length of such a string of grains is
\begin{equation}
L=n (2 r_0 - \delta _0)+\delta _0 - 2 r_0.
\label{length_spring_set}
\end{equation}
Since the spring constant for a set of springs is simply $k_L=k/n$, the sound
velocity can be written as
\begin{equation}
c_\mathrm{s}=\sqrt{\frac{k_L L}{\rho _\mathrm{l}}}=\sqrt{\frac{k L^2/n}{n
m_0}}=\frac{L}{n} \sqrt{\frac{k}{4/3 \pi r_0^3 \rho _0}}.
\label{eq::soundspeed:pre}
\end{equation}
Substituting eqs.~\ref{spring_const} and \ref{length_spring_set} we get
\begin{equation}
c_\mathrm{s}=\biggl(1-\frac{1}{n}\biggr)(2 r_0 - \delta _0) 
\sqrt{\frac{(9\pi \gamma R^2 E^{\star 2})^{1/3} 6/5} {4 \pi r_0^3 \rho _0/3}}.
\label{eq::sounspeed:fin}
\end{equation}

Now we can try to see how the sound velocity in such system depends
on the size of a single monomer.
\begin{equation}
c_\mathrm{s} \propto r_0 \sqrt{\frac{r_0^{2/3}}{r_0^3}}=r_0^{-1/6}.
\end{equation}
This means that to double the speed of sound we need to use $64$ times smaller
grains. Therefore, the sound speed in this case is a very weak function of
monomer size.

\begin{table}
\caption{Material properties used in this work for silica.}
\label{tab:material_properties}
\centering
\begin{tabular}{c c c c c}
\hline\hline
$E^{*}$[dyn/cm$^2$] & $\gamma$ [erg/cm$^2$] & $\rho_0$ [g/cm$^3$]\\
\hline
$2.78 \cdot 10^{11}$ & $25.0$ & $2.65$ \\
\hline
\end{tabular}
\end{table}

The sound speed of an infinitely long chain of $0.6$~$\mu$m silica grains is
$c_\mathrm{s}=513$~m/s. If we apply our findings to a chain of $50$,
$0.6$~micron sized, silica grains, the sound velocity turns out to be
$c_\mathrm{s}=503$~m/s. We can compare it now to the previous results obtained
in research of granular medium composed of mm sized and bigger
grains. In all our simulations and analytical calculations we used
  material properties as specified in tab.~\ref{tab:material_properties}.

\citet{1999PhRvE..59.3202H} developed a model of macroscopic grains to study the
propagation of sound in a granular chain. The centimeter sized grains they use
do not interact in this case via attractive surface forces. The signal is
transported due to the Hertzian stress that arises as an effect of overlap of
grains in contact. They also apply the spring theory but with an arbitrarily
chosen spring constant $k$.  For values of $k$ in the range between $k=10^6$~N/m
and $k=10^8$~N/m, and mass density of $1.9\times 10^3$~kg/m$^3$, the sound
velocity they derive is in a range of $300$~m/s to $3000$~m/s. Similar results
were obtained by \citet{2006JSMTE..07..023M}. A sound speed $c_\mathrm{s}\approx
200$~m/s was found for closely packed grains. In this case monomer radius was
$1$~mm, density $2\times 10^3$~kg/m$^3$, and the spring constant $k=10^5$~N/m.

\subsubsection{Numerical experiment}
In order to verify our findings we performed a numerical experiment. We prepared
a linear chain of 50 monomers.  The first one in line was slightly displaced
from it's equilibrium position. When the simulation started the grain started to
move in order to reach it's equilibrium and the second grain was disturbed. Such
motion was propagating until it finally reached the last grain. The total
distance traveled by the density wave was $99(r_0 - \delta_0)$. For $50$ silica
monomers with radii $r_0=0.6$~$\mu$m the travel time took $t=1.16\,10^{-7}$~s,
which results in the sound speed of $c_\mathrm{s}=512$~m/s. We can now compare
the value with the theoretically derived sound speed $c_\mathrm{s}=503$~m/s. The
difference between the theoretical velocity and the one that was obtained
numerically is only about $3\%$. Moreover in the simulation the first grain is
displaced by a finite distance meaning that the assumption of low displacements
may not be entirely correct. The displacement of about $0.5 \, \delta_0$ is
relatively large. We performed a series of simulations with stronger
perturbations and the data obtained shows that the sound velocity increases as
the perturbation strength increases.

\subsubsection{Porous aggregates}
Since a linear chain of monomers is a special case, we will discuss now the more
general agglomerates of irregular shape. The sound speed in such a system is
affected by several things. Firstly, the path length that a signal has to travel
\begin{figure}[!h]
\resizebox{\hsize}{!}{\includegraphics{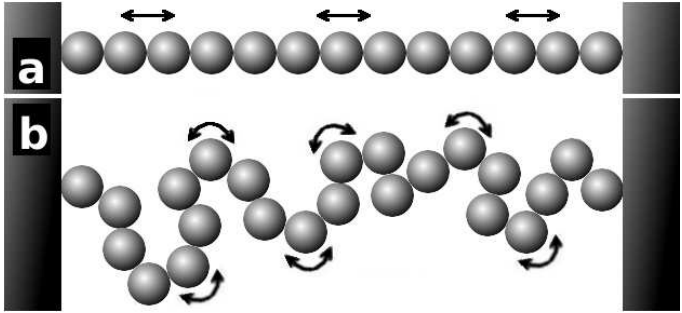}}
\caption{Sketch of a sound wave propagation in a linear aggregate (a) and in a
porous aggregate (b).}
\label{fig::sketch_cs}
\end{figure}
is not a straight line and thus with a given speed, longer distances result in
lower \emph{effective} sound speed. For a small RBD aggregate (approximately 300
monomers), the \emph{effective distance} was larger by only a factor of
$1.5$. Thus structure of an aggregate has a limited impact on the sound velocity
by increasing the path length. The second factor that may have an effect is the
tangential force. In a linear aggregate, the signal is passed forward due to the
presence of the vertical force, and the tangential force has no effect. However,
in irregular aggregates grains interact also via rolling and sliding. This
might lower the contribution of the vertical force and in this way change the
sound speed. Fig.~\ref{fig::sketch_cs} shows a sketch of how a sound wave
propagates in different aggregates. Indeed linear aggregates involve only
stronger, vertical forces and thus the signal travels faster. Irregular particles
also make use of the tangential, rolling force.

In order to estimate the sound speed due to the rolling friction, we use again
the spring theory. \citet{1995PhMA...72..783D} gives the recipe for the rolling
friction to be
\begin{equation}
F_\mathrm{roll}=6 \pi \gamma \xi,
\label{eq::rollingfriction}
\end{equation}
where the $\xi$ is a displacement from the equilibrium position. This force can
be then expressed as 
\begin{equation}
F_\mathrm{roll}=k \xi,
\label{eq::rollingsoundspeed}
\end{equation}
with $k=6 \pi \gamma$ -- a spring constant. We then apply this in
eq.~\ref{eq::soundspeed:pre} to get 
\begin{equation}
c_\mathrm{s}=\biggl(1-\frac{1}{n}\biggr)(2r_0-\delta_0)\sqrt{\frac{6 \pi \gamma}{4/3 \pi r_0^3 \rho
_0}}.
\label{eq::soundspeed:rol:fin}
\end{equation}
The dependence of the sound speed on a grain size, $c_\mathrm{s} \propto
r_0^{-1/2}$, is in this case much stronger than in the case of the vertical force
($c_\mathrm{s} \propto r_0^{-1/6}$).

When we apply eq.~\ref{eq::soundspeed:rol:fin} to the $1.2 \, \mu$m silica
grains we get sound speed $c_\mathrm{s}=16.5$ m/s. This is a factor of $30$
lower than the velocity derived for the linear chain. This suggests that the
sound velocity in a porous aggregate might be significantly different from the
speed derived for a linear chain of grains.

\citet{2004Icar..167..431S} derived compressive and tensile strengths dependence
on density, based on experimental data. That relations were later used to
develop an SPH model of large, mm sized and larger aggregates collisions. The
sound speed in an aggregate made of $0.1\,\mu$m ice monomers was calculated to
be $c_\mathrm{s}=\sqrt{E/\rho}$. He used values for the Young's modulus and the
density to be $E=6\times10^5$~Pa and $\rho=0.1$~g/cm$^3$, respectively. The
resulting sound speed is then $c_{\mathrm{s}}=77.5$~m/s.

When we apply our formula for the sound speed to $100$ aligned $0.1\,\mu$m ice
grains, we get sound speed $c_\mathrm{s}=885$ m/s. This is almost an order of
magnitude higher. We have to keep in mind however that the speed in a linear
chain may be considerably different to the one in a real porous aggregate. In
the rolling case, the theoretical sound speed is also high with
$c_\mathrm{s}=250$ m/s.

\citet{2007thesis.book..tu-bs..T} performed a lab experiment, where he
measured the sound speed in a RBD aggregate with filling factor $\phi=0.15$ and
$1.5\,\mu$m silica monomers. He hit the dust cake from below and measured the
response at the surface with a force sensor. The measured velocity was
$c_\mathrm{s}=30 \pm 4$ m/s, very much consistent with our results when the signal is
assumed to be transferred through the rolling degree of freedom.

\begin{figure}[!h]
\resizebox{\hsize}{!}{\includegraphics{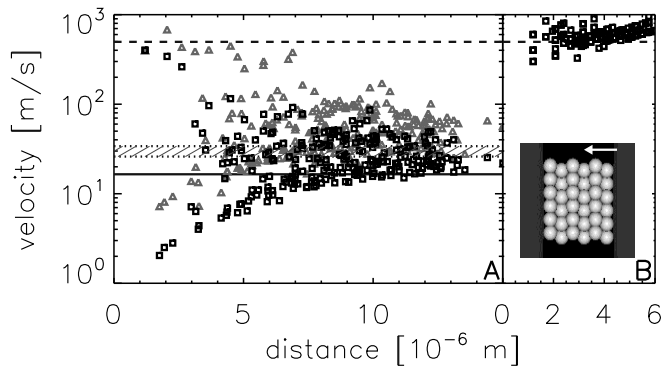}}
\caption{Sound velocity in m/s versus distance from the plane that hits the
  aggregate. Squares indicate applied force of $5\,10^{-8}$~N and triangles
  $5\,10^{-6}$~N. The left panel (a) shows results of sound speed determination
  in a RBD aggregate, while the right one (b) in a CCP aggregate. The inset
  image shows the CCP aggregate placed in between two planes. The lines indicate
  sound velocities determined theoretically using vertical force in a linear
  chain of monomers (dashed line) and using rolling friction (solid line). The
  Dashed area limited by dotted lines indicates the sound speed range determined
  experimentally by \citet{2007thesis.book..tu-bs..T}}
\label{fig::soundspeed}
\end{figure}

In order to numerically derive the sound speed and further test our model we
performed a simulation of an RBD dust cake. The cake was being pushed from one
side and we determined the response time of different particles in the
aggregate. We combined the positions of the particles with their response time,
which results in an average sound speed in the dust
cake. Fig~\ref{fig::soundspeed}a shows the result of our simulations. For each
monomer in the dust cake we plotted a corresponding sound speed. Initially large
spread in the data shows that inside the cake very linear structures are
present. This leads to a few particles with very large sound speed. At larger
distances, however, the range of sound speeds is much lower and shows that a
signal is transported mainly via rolling degree of freedom. The average sound
speed that is determined at the far end of the dust cake is only a factor of
about $1.5$ larger than the sound velocity in the rolling degree of freedom.

We applied two different forces to the cake and later determined mean sound
speeds in the cake for both cases. When a stronger force of $5 \times 10^{-6}$ N
was applied to the cake the sound speed reached $c_\mathrm{s} \approx 60$~m/s,
while with the lower force of $5 \times 10^{-8}$ N the sound velocity was
$c_\mathrm{s}\approx 20$~m/s. For one monomer the limiting sound velocity, as
calculated for linear chain of monomers was overcome. This, however, happened in
the case of larger force, and thus the low displacement approximation was most
likely violated leading to larger sound speeds.

The simulations show that the sound propagation is dominated by the rolling
friction. In order to verify this we run another simulation to determine the
sound speed in the aggregate with monomers arranged according to the cubic close
packing (CCP) because this arrangement disables any rolling motion. The inset in
fig.~\ref{fig::soundspeed}b shows the aggregate. We applied a force on one side
of the agglomerate and determined the response time of different particles. Data
shown in fig.~\ref{fig::soundspeed}b shows that indeed closely packed aggregates
are characterized by much higher velocities than the porous ones. The monomers
in each layer received the signal at a different time because the compressing
planes were simulated by big grains and thus central particles were hit first
and they forwarded the signal. Thus particles at the edges of the aggregate
responded later than the ones in the center of each layer. The velocity
$c_\mathrm{s} \approx 600$~m/s shows that indeed sound speed in a porous medium
strongly depends on porosity because of the forces involved in the transport of
the signal and the longer path for the signal to travel in more porous
aggregates. Using a constant sound speed in SPH simulations is bound to lead to
spectacularly wrong results.

\section{Conclusions}
We have modified the original code by \citet{1997ApJ...480..647D} and
\citet{2002Icar..157..173D}.  We have shown that a numerical N-particle method
for studying the properties of aggregates can be calibrated to experimental
results by including the flattening of small asperities on the surface of the
grains, and by using critical displacements for rolling of grains consistent
with measured values.  The new code reproduces the measured critical velocities
for sticking very well.  We would like to emphasize that there is currently no
proof for plastic deformation actually happening at the grain surfaces, but as a
model it works very well.

We went on to measure the compressive strength of the aggregates and compared
the results with experiments.  We get very similar results with a small offset
in porosity which is most likely due to the relatively small aggregates used in
our simulations.  Finally, we computed the sound velocity in an
adhesion-dominated porous material and showed that this leads to very
interesting results.  Three very different velocities play a role in such a
medium, and the different velocities are separate by factors of the order of 10.
The fastest speed is the bulk material sound speed which only plays a role after
a body has been molten and re-hardened.  The second speed, a factor of $\sim$10
lower is the speed at which a signal is transported in a longitudinal wave in a
linear chain.  This speed applies either in a perfectly linear chain, or in an
aggregate that has been compacted sufficiently so that rolling of grains is no
longer possible.  Finally, the slowest speed is the one transmitted by rolling
forces in a non-straight chain of grains.  A small decrease stems from the
longer path the sound has to take in a porous aggregate.  By far the largest
fraction stems from the weak forces in the rolling degree of freedom.
Experiments show that this is indeed the dominant speed in porous aggregates.
However, our results show that the sound speed should be a very steep function
of density once a significant number of monomers has 3 or more contacts with their
neighbors. It is to be expected that SPH approaches to model the properties of
dust aggregates \citep{2004Icar..167..431S,2007A&A...470..733S} will fail
strongly if these effects are not taken into account properly.

In summary, we conclude that we do now have a working model of dust
aggregates that can be applied in parameter studies.

\begin{acknowledgements}
This work was supported by the Nederlandse Organisatie voor Wetenschapelijk
Onderzoek, Grant 614.000.309. We thank J\"urgen Blum for useful discussions and
hospitality during several visits. We also acknowledge a financial support of
Leids Kerkhoven-Bosscha Fonds.
\end{acknowledgements}

\bibliographystyle{aa} 
\bibliography{draft}
\end{document}